\documentclass[sigconf,nonacm]{acmart}

\usepackage{tikz}
\usetikzlibrary{arrows.meta, positioning, calc}
\usepackage{graphicx}
\usepackage{dsfont}
\usepackage{subcaption}
\usepackage{booktabs}
\usepackage{tabularx}
\usepackage[nolist]{acronym}
\usepackage{xcolor}
\usepackage{mathtools}
\usepackage{hyperref}
\usepackage{url}
\usepackage{algorithm}
\usepackage{algorithmic}
 \usepackage{tikz}
 \usepackage{booktabs}
\usepackage{tabularx}
\usepackage{multirow}
\usepackage{makecell}
\AtBeginDocument{%
  }

\begin{document}
\title{Evolution of Financial AI: From Statistical Models to Agentic AI}


 \author{Saba Asaad}
 \email{saba.asaad@utoronto.ca}
 \affiliation{%
   \institution{University of Toronto}
   \city{Toronto}
   \country{Canada}
 }
 \author{Ali Bereyhi}
 \email{ali.bereyhi@utoronto.ca}
 \affiliation{%
   \institution{University of Toronto}
   \city{Toronto}
   \country{Canada}
 }

\renewcommand{\shortauthors}{Asaad et al.}
\begin{acronym}
\acro{pca}[PCA]{principal component analysis}
\acro{ai}[AI]{artificial intelligence}
\acro{ml}[ML]{machine learning}
\acro{dl}[DL]{deep learning}
\acro{svm}[SVM]{support vector machine}
\acro{fnn}[FNN]{feedforward neural network}
\acro{cnn}[CNN]{convolutional neural network}
\acro{rnn}[RNN]{recurrent neural network}
\acro{gnn}[GNN]{graph neural network}
\acro{mlp}[MLP]{multilayer perceptron}
\acro{lstm}[LSTM]{long short-term memory}
\acro{gru}[GRU]{gated recurrent unit}
\acro{rag}[RAG]{retrieval-augmented generation}
\acro{llm}[LLM]{large language model}
\end{acronym}

\begin{abstract}

Artificial intelligence (AI) in finance has evolved from statistical learning and machine learning for structured prediction, through deep learning and foundation models for representation learning, language understanding, and reasoning, to retrieval-augmented generation for knowledge-grounded reasoning and agentic AI for autonomous planning and execution. In this paper, we examine how researchers and practitioners can select and apply AI methods in finance. We first show that the major paradigms of financial AI address fundamentally different classes of problems, rather than forming a strict hierarchy of capability. We then review the current state of research on AI in finance and trace how these paradigms have reshaped financial data, financial tasks, decision support, and human-AI collaboration.
Building on this, we propose practical guidelines for matching AI paradigms to task characteristics, and show that more advanced models, such as foundation models or autonomous agents, are not universally superior. We conclude by outlining open challenges for researchers and practitioners working toward reliable financial AI.
\end{abstract}

\maketitle

\section{Introduction}
Consider a financial institution deciding how to automate three very different tasks: scoring a loan applicant's creditworthiness, answering a client's question about a new regulatory disclosure, and conducting a multi-step due-diligence review before an acquisition. A growing tendency in financial \ac{ai} is to approach such different problems with the same highly capable technology, a \ac{llm} or an autonomous agent, even when the task does not require its full capabilities. Yet, a gradient-boosted tree trained on historical repayment data may outperform an \ac{llm} at the first task, a retrieval-augmented system is better suited to the second, and the third may require an agentic workflow capable of coordinating multiple analytical steps. This mismatch between \textit{task} and \textit{technology} is what motivates this study.

Over the past three decades, financial AI has progressed through several major paradigms, from statistical and \ac{ml} methods to \ac{dl}, foundation models, retrieval-augmented systems, and agentic AI. Each new paradigm emerged to address the specific limitations of its predecessor, progressively expanding what financial \ac{ai} can do, from prediction and representation learning to natural language understanding, knowledge foundation and autonomous task execution. However, financial \ac{ai} research and practice often treat these paradigms in isolation, studying individual methods or applications rather than situating them within this broader evolution. As a result, the field is frequently viewed through a hierarchy in which newer paradigms simply supersede older ones, rather than as a set of complementary tools suited to different classes of financial tasks. This view makes it difficult to determine which paradigm fits a given task and encourages a tendency to reach for the most advanced methods, even where a simpler, cheaper, and more reliable approach would do.

This survey addresses this gap by providing a unified perspective on the evolution of financial \ac{ai} and its implications for financial decision-making. Our contributions are as follows:
\begin{enumerate}
    \item \textbf{A unified evolutionary account.} We trace the evolution of financial \ac{ai} from statistical learning to agentic AI, highlighting the limitations that motivated each paradigm transition.
    \item \textbf{A capability-oriented perspective.} We examine how successive \ac{ai} paradigms have transformed financial information processing, intelligent financial tasks, decision support, and human-\ac{ai} collaboration.
    \item \textbf{A practical selection framework.} We propose guidelines for matching \ac{ai} paradigms to the characteristics of financial tasks, showing that \textit{more advanced paradigms are not universally preferable.}
    \item \textbf{Open challenges.} We identify the key technical and organizational challenges that must be addressed to enable reliable, trustworthy, and responsible financial \ac{ai}.
\end{enumerate}

\section{Evolution of AI in Finance}
The development of financial \ac{ai} can be understood as a sequence of responses to changing data, computational capabilities, and task requirements \cite{du2025natural,pazouki2025artificial}. As financial problems expanded beyond structured prediction to include unstructured information, reasoning, knowledge integration, and multi-step decision processes, new AI paradigms emerged to address capabilities that earlier approaches lacked. These paradigms remain complementary, with each suited to different types of financial problems. Figure~\ref{fig:ai-finance-timeline} summarizes the major transitions examined in this section.

\subsection{Statistical Learning: Predictive Modeling}
The foundations of modern financial AI can be traced to statistical learning methods, whose primary objective is to model financial data for prediction and inference. These methods estimate the statistical relationship between observed variables and a target outcomes, enabling predictions for previously unseen observations. Depending on the availability of labeled data, learning methods can be broadly categorized into supervised and unsupervised learning \cite{hastie2009elements}. %
Supervised learning operates on labeled datasets, where each observation consists of input features and a corresponding target variable. It has been widely applied to financial problems such as credit scoring, prediction, risk assessment, and asset pricing. Early statistical methods relied primarily on linear and logistic regression, due to their simplicity, interpretability, and theoretical foundations \cite{bishop2006pattern}. However, these models assume relatively simple relationships between variables and often struggle to capture nonlinear relationships and interactions among financial variables. Their performance can also degrade in high-dimensional settings, where overfitting becomes a major concern. To improve generalization while preserving interpretability, regularization techniques such as Ridge and LASSO regression were introduced, enabling more robust statistical models for financial prediction \cite{hoerl1970ridge, tibshirani1996regression}.
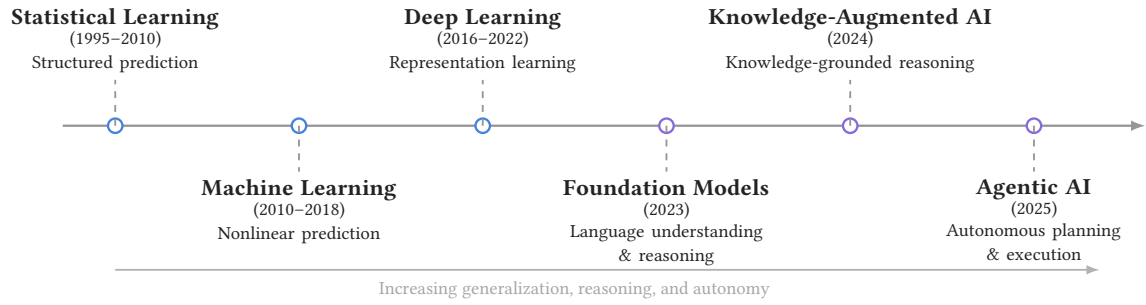
\begin{figure*}[t]
\label{fig:ai-finance-timeline}
\centering
\resizebox{0.85\textwidth}{!}{%
\begin{tikzpicture}[font=\sffamily]

\definecolor{clblue}{RGB}{70,130,220}
\definecolor{clpurple}{RGB}{135,105,220}
\definecolor{timeline}{RGB}{150,150,150}
\definecolor{textgray}{RGB}{35,35,35}

\draw[timeline, line width=1.0pt, -{Latex[length=2.2mm]}] (0,0) -- (16.5,0);

\tikzset{
  classical/.style={
    circle,
    draw=clblue,
    fill=white,
    line width=1.0pt,
    minimum size=2.2mm,
    inner sep=0pt
  },
  generative/.style={
    circle,
    draw=clpurple,
    fill=white,
    line width=1.0pt,
    minimum size=2.2mm,
    inner sep=0pt
  },
  topbox/.style={
    align=center,
    text=textgray,
    font=\normalsize,
    text width=3.2cm
  },
  bottombox/.style={
    align=center,
    text=textgray,
    font=\normalsize,
    text width=3.2cm
  }
}

\coordinate (p1) at (0.8,0);
\coordinate (p2) at (3.6,0);
\coordinate (p3) at (6.4,0);
\coordinate (p4) at (9.2,0);
\coordinate (p5) at (12.0,0);
\coordinate (p6) at (14.8,0);
\node[classical] at (p1) {};
\node[classical] at (p2) {};
\node[classical] at (p3) {};
\node[generative] at (p4) {};
\node[generative] at (p5) {};
\node[generative] at (p6) {};

\draw[timeline, dashed, line width=0.8pt] (p1) -- ++(0,0.7);
\draw[timeline, dashed, line width=0.8pt] (p2) -- ++(0,-.7);
\draw[timeline, dashed, line width=0.8pt] (p3) -- ++(0,0.7);
\draw[timeline, dashed, line width=0.8pt] (p4) -- ++(0,-0.7);
\draw[timeline, dashed, line width=0.8pt] (p5) -- ++(0,0.7);
\draw[timeline, dashed, line width=0.8pt] (p6) -- ++(0,-0.7);

\node[topbox, above=0.7cm of p1] {%
\textbf{\large Statistical Learning}\\[-0.8mm]
{\small (1995--2010)}\\[-0.2mm]
\small Structured prediction
};

\node[bottombox, below=0.7cm of p2] {%
\textbf{\large Machine Learning}\\[-0.8mm]
{\small (2010--2018)}\\[-0.2mm]
\small Nonlinear prediction
};

\node[topbox, above=0.7cm of p3] {%
\textbf{\large Deep Learning}\\[-0.8mm]
{\small (2016--2022)}\\[-0.2mm]
\small Representation learning
};

\node[bottombox, below=0.7cm of p4] {%
\textbf{\large Foundation Models}\\[-0.8mm]
{\small (2023)}\\[-0.2mm]
\small Language understanding \\[-0.2mm]
\small \& reasoning
};

\node[topbox, text width=4.8cm, above=0.7cm of p5] {%
\textbf{\large Knowledge-Augmented AI}\\[-0.8mm]
{\small (2024)}\\[-0.2mm]
\small Knowledge-grounded reasoning
};


\node[bottombox, below=0.7cm of p6] {%
\textbf{\large Agentic AI}\\[-0.8mm]
{\small (2025)}\\[-0.2mm]
\small Autonomous planning\\[-0.2mm]
\small \& execution
};

\draw[gray!60, line width=0.8pt,-{Latex[length=2mm]}]
(0.8,-2.2) -- (15.8,-2.2);

\node[gray!70,font=\small]
at (7.8,-2.5)
{Increasing generalization, reasoning, and autonomy};

\end{tikzpicture}%
}
\caption{Evolution of AI paradigms in finance. Financial AI has shifted from statistical prediction toward autonomous decision-making systems. Each paradigm addresses key limitations of its predecessor while complementing, rather than replacing, earlier approaches.}
\label{fig:ai-finance-timeline}
\end{figure*}

In contrast, unsupervised learning aims to uncover hidden structures in unlabeled data without predefined target variables \cite{bishop2006pattern}. Common approaches include clustering techniques for customer segmentation and anomaly detection, as well as dimensionality reduction methods such as \ac{pca}, which compress high-dimensional financial data into a smaller set of informative features \cite{jolliffe2002principal}. These methods have been widely applied to financial applications, including portfolio construction, risk management, fraud detection, and customer behavior analysis.

Despite their widespread adoption, statistical learning methods have several limitations in modern financial applications. Most rely on structured numerical data and often require extensive feature engineering to capture complex relationships. Their ability to process unstructured information, such as financial reports, news articles, earnings calls, and social media, remains limited while many methods struggle to model the nonlinear and high-dimensional relationships characteristic of modern financial data. These limitations motivated the transition toward \ac{ml}, where models could learn increasingly complex relationships directly from data with fewer predefined modeling assumptions \cite{pazouki2025artificial}.
\subsection{ML and DL: Learning Complex Patterns}
\Ac{ml} marked a significant advancement in financial \ac{ai} by enabling models to automatically discover predictive patterns from the data. This capability substantially improved performance on complex problems that involve large, nonlinear, and high-dimensional datasets, leading to widespread adoption across prediction and classification in finance \cite{dixon2020machine}.
Representative \ac{ml} methods include decision trees, random forests, boosting algorithms, and \ac{svm}. %
Among these, tree-based methods have been particularly successful for structured financial data because they naturally capture nonlinear relationships and interactions among variables while remaining relatively interpretable \cite{steinberg2009cart}.  Ensemble methods such as random forests and boosting further improve robustness and predictive performance by combining multiple decision trees, making them among the most widely adopted techniques for financial prediction and risk assessment \cite{breiman2001random, chen2016xgboost}. 
\acp{svm} provide an alternative approach that is particularly effective for high-dimensional classification problems \cite{cortes1995support}. These methods have been widely applied to credit scoring, fraud detection, bankruptcy prediction, risk assessment, and asset price forecasting, due to their stability and flexibility. The rapid growth of financial data further accelerated the adoption of \ac{dl}, which extends \ac{ml} through representation learning by automatically extracting hierarchical features directly from raw data with minimal feature engineering \cite{goodfellow2016deep}. Its success in finance has been driven by its ability to model complex nonlinear relationships, capture long-term temporal dependencies in sequential data, and leverage large-scale heterogeneous datasets. As a result, a set of neural architectures has emerged, each designed to address characteristics of financial data and applications \cite{huang2020deep}.

\Ac{dl} methods can be categorized according to the type of data they are designed to process. In finance, these include structured numerical data, financial time series, unstructured text, and relational data describing interactions among financial entities. Representative architectures include fully-connected \acp{fnn}, \acp{cnn}, \acp{rnn}, transformers, and \acp{gnn} \cite{goodfellow2016deep, vaswani2017attention, kipf2016semi}. \acp{fnn}, also referred to as \acp{mlp}, are the simplest class of neural models and have been widely used in finance for nonlinear prediction and classification tasks such as credit scoring and customer segmentation \cite{goodfellow2016deep}.
\acp{cnn} effectively capture local spatial patterns through convolutional operations, making them suitable for applications such as fraud detection and financial time-series analysis \cite{lecun1998gradient}. \acp{rnn}, particularly the \ac{lstm} and \ac{gru} architectures, model temporal dependencies in sequential data and have been extensively applied to stock price prediction, volatility forecasting, and algorithmic trading \cite{hochreiter1997long}. More recently, transformers, through the self-attention mechanism, have demonstrated promising enhancements in capturing long-range dependencies, substantially improving the analysis of sequential and textual data and introducing the notion of modern \textit{foundation models} \cite{vaswani2017attention}. In parallel, \acp{gnn} have emerged as powerful models for relational financial data, including transaction networks, payment systems, and corporate ownership structures \cite{cho2014learning}.

Despite their remarkable predictive performance, \ac{dl} models present several challenges in financial applications. They typically require large labeled datasets, operate as black-box models with limited interpretability, and remain largely task-specific, limiting their ability to perform complex reasoning or integrate heterogeneous financial knowledge. These limitations motivated the emergence of foundation models, which extend \ac{dl} through large-scale pretraining to enable natural language understanding and generalization across diverse  tasks\cite{bommasani2021opportunities}.

\subsection{Foundation Models: Toward Understanding}
Foundation models represent a fundamental shift in financial \ac{ai} from task-specific predictive models to general-purpose models capable of natural language understanding, reasoning, and generation \cite{bommasani2021opportunities}. Unlike conventional \ac{ml} and \ac{dl} models that are typically developed for a single downstream task, foundation models are pretrained on massive corpora and subsequently adapted to diverse financial applications through prompting or fine-tuning. This paradigm expands financial \ac{ai} beyond prediction, enabling language understanding, reasoning, summarization, question answering, and other analytical tasks across a broad range of documents and data sources \cite{li2023large}.

The emergence of foundation models was enabled by the transformer architecture introduced by Vaswani et al. in 2017 \cite{vaswani2017attention}.  Unlike \acp{rnn}, transformers employ self-attention mechanism that captures long-range dependencies while enabling efficient parallel training on large-scale datasets. Based on the generic transformer architecture, foundation models can be broadly categorized into encoder-only, encoder-decoder, and decoder-only models, each supporting different classes of financial applications. Encoder-only models, such as BERT \cite{devlin2019bert}, are used primarily for language understanding tasks, including financial sentiment analysis, document classification, and information extraction. Encoder-decoder models, such as T5 \cite{raffel2020exploring} and BART \cite{lewis2020bart}, support both language understanding and generation, making them well suited for document summarization and question answering. Decoder-only models, represented by GPT-style architectures, excel at autoregressive text generation and have demonstrated strong capabilities in financial reasoning and decision support \cite{brown2020language}. Despite advances, general-purpose foundation models lack domain-specific financial knowledge and current market information, limiting their effectiveness in financial applications.

Financial text often involves specialized terminology, domain-specific context, and implicit numerical reasoning that may not be fully captured by models pretrained on general corpora. These challenge have motivated the development of \textit{finance-specific foundation models.} Early efforts adapted models such as BERT to financial tasks, leading to FinBERT \cite{araci2019finbert}, which demonstrated strong performance in financial sentiment analysis and named entity recognition. More recently, larger finance-specific models such as FLANG \cite{shah2022flue} and BloombergGPT \cite{wu2023bloomberggpt} have been pretrained on substantially larger financial corpora, improving performance over a broader range of applications, including financial question answering, document understanding, summarization, and information extraction. These domain-adapted models enhance the capabilities of financial \ac{ai} by improving the analysis of unstructured financial information and supporting advanced decision-support applications.

Despite their remarkable capabilities, foundation models remain limited by the knowledge acquired during pretraining. As a result, their responses may become outdated or factually inaccurate in rapidly evolving financial environments where regulations, market conditions, and corporate information continuously change. In addition, numerical reasoning, factual consistency, and explainability remain challenging, limiting their reliability in high-stakes financial decision-making. These limitations motivated the development of knowledge-augmented AI systems, which integrate foundation models with external knowledge sources to provide more accurate, up-to-date, and trustworthy financial decision support \cite{lewis2020retrieval}.

\subsection{Knowledge-Augmented AI}

A fundamental limitation of foundation models is that their knowledge remains limited to the information acquired during pretraining \cite{bommasani2021opportunities}. While techniques such as fine-tuning enable adaptation to domain-specific tasks, it embeds knowledge within model parameters and therefore cannot continuously incorporate new information. As a result, foundation models may generate outdated or incomplete responses in rapidly evolving financial environments where market conditions, regulations, and corporate information continually change. 
Knowledge-augmented AI complements foundation models with external knowledge retrieved during inference, enabling more accurate and domain-aware reasoning rather than relying solely on pretrained knowledge \cite{lewis2020retrieval}.

The \ac{rag} is the most adapted framework for implementing knowledge-augmented AI. It combines a pretrained foundation model with an external retrieval system, allowing relevant information to be retrieved from external knowledge repositories and incorporated into the generation process during inference \cite{lewis2020retrieval}.
By grounding responses in current and domain-specific financial information, \ac{rag} improves factual accuracy, reduces hallucinations, and enables reasoning over information unavailable during pretraining. This capability is particularly valuable in finance, where market conditions, regulations, and corporate disclosures evolve continuously. As a result, \ac{rag} supports applications such as financial question answering, document analysis, compliance assistance, and investment research by combining the reasoning capabilities of foundation models with real-time market data, regulatory documents, financial reports, and organization-specific knowledge \cite{iaroshev2024evaluating}.

Beyond retrieval, knowledge augmentation can also be achieved through structured knowledge representations such as knowledge graphs \cite{hogan2021knowledge}. Unlike retrieval-based methods that dynamically access external documents, knowledge graphs provide a structured representation of financial knowledge, enabling foundation models to reason over explicit relationships among financial entities. A typical knowledge graph consists of nodes representing entities and edges representing the relationships between them. In finance, entities such as companies, shareholders, transactions, and regulatory bodies form rich and highly interconnected relationships that are difficult to capture using unstructured text alone. By explicitly modeling these relationships, knowledge graphs enable foundation models to exploit structured financial knowledge, supporting applications including financial question answering, fraud detection, investment analysis, and regulatory compliance \cite{ji2021survey}.

Despite these advantages, knowledge-augmented AI systems introduce new challenges in financial applications. Their effectiveness depends critically on the quality, coverage, and timeliness of external knowledge sources \cite{patel2024survey}. Furthermore, effectively combining structured knowledge with the implicit knowledge learned by foundation models remains challenging, particularly for numerical reasoning, factual consistency, and explainability. Scalability, latency, data privacy, and governance also become increasingly important as these systems interact with large-scale and continuously evolving financial information.
These challenges have motivated the development of more autonomous AI systems that combine reasoning, retrieval, memory, and tool use to support increasingly reliable and intelligent financial decision-making.

\subsection{Agentic AI: Toward Autonomous Finance}
Agentic AI represents the latest stage in the evolution of financial AI, extending foundation models and knowledge-augmented AI from information understanding and decision support toward autonomous financial systems \cite{acharya2025agentic}. Unlike foundation models, which primarily assist with individual analytical tasks, agentic AI is designed to coordinate the multiple steps required to accomplish complex financial objectives by integrating reasoning, external knowledge, memory, planning, and specialized analytical tools within a unified workflow. As a result, financial AI evolves from supporting isolated analytical tasks to coordinating end-to-end workflows such as investment research, compliance assessment, credit evaluation, and financial reporting with reduced human intervention.

In financial applications, agentic \ac{ai} differs from the earlier \ac{ai} paradigms in three important respects. First, it supports \textit{end-to-end workflows} by decomposing complex objectives, such as investment analysis, credit assessment, or compliance investigations into coordinated analytical steps. Second, it performs \textit{contextual reasoning} by continuously incorporating evolving financial information, including market data, corporate disclosures, regulatory updates, and organizational knowledge. Third, it \textit{interacts} directly with financial infrastructure, including analytical software, enterprise databases, trading platforms, and risk management systems. This enables agents to execute analytical tasks rather than recommend actions.

These capabilities enable agentic \ac{ai} to support a broad range of financial activities that require coordinating multiple analytical steps. In investment research, agents can retrieve financial information, analyze corporate disclosures, and prepare analytical reports. In financial market analysis and algorithmic trading, they can continuously monitor market conditions and adapt trading strategies in response to new information. In risk management and compliance, agents can perform stress testing, monitor regulatory updates, and assist with audit and reporting processes. More broadly, agentic AI shifts financial AI from generating recommendations to executing end-to-end workflows under human supervision \cite{aldridge2026agentic,acharya2025agentic}.

Despite its significant potential, agentic AI introduces important challenges for financial applications. The autonomous and multi-step nature of agentic decision-making \textit{reduces transparency}, making explainability, accountability, and regulatory compliance more difficult. In addition, ensuring reliable performance as market conditions and financial information continuously change remains a significant challenge. As autonomous agents become more widely adopted across financial institutions, their interactions can also create unexpected behaviors and increase systemic risk. Addressing these challenges is essential before agentic AI can be widely deployed in high-stakes financial applications.


\section{From Prediction to Decision Support}
The previous section traced the evolution of AI in finance from statistical learning to agentic AI. This section shifts the focus from AI paradigms to their impact on financial systems. We examine how advances in AI have transformed financial information, intelligent financial tasks, decision support, and human-AI collaboration.

\subsection{Data Modeling vs. Multimodal Information}
Traditional financial AI primarily relied on structured data, where information is represented in predefined formats such as historical prices, trading volumes, financial statements,  consensus estimates, and macroeconomic indicators \cite{pazouki2025artificial}. These numerical data have long served as the foundation of quantitative finance, enabling statistical and \ac{ml} models to perform forecasting, classification, and risk analysis \cite{hastie2009elements, bishop2006pattern}. However, modern financial decision-making increasingly relies on unstructured and multimodal information that cannot be naturally represented in predefined tabular formats. Today, financial information extends far beyond numerical records to include complementary textual, audio, and visual sources. Textual information—including financial news, analyst reports,  corporate disclosures, and social media captures qualitative information about firms, markets, and investor sentiment \cite{cao2022ai}. Audio sources, such as earnings calls, investor presentations, and conference discussions, provide behavioral cues through speech characteristics and management tone \cite{hajek2023speech}. Visual information, including charts, scanned documents, and satellite imagery, complements numerical and textual data by supporting document understanding and revealing patterns related to market activity and economic conditions \cite{duane2025review}.

The growing diversity of financial information required corresponding advances in AI models capable of processing heterogeneous data. Earlier \ac{ml} methods, such as \acp{svm}, random forests, and gradient boosting models, primarily relied on handcrafted features and modality-specific representations \cite{cortes1995support,breiman2001random}.
Although these approaches could incorporate textual information through manually engineered features such as bag-of-words representations or sentiment lexicons, they lacked the ability to learn rich semantic features directly from raw financial data or to integrate heterogeneous information within a unified model. \Ac{dl} introduced representation learning through architectures such as \acp{cnn} for visual data, \acp{rnn} for time series and text, and later transformers, enabling models to automatically learn informative features from raw financial data \cite{goodfellow2016deep,vaswani2017attention}.
However, these models generally remained specialized for individual modalities or tasks, requiring separate architectures for text, speech, images, and numerical data. Foundation models leverage large-scale pretraining and self-attention to learn shared multimodal representations, enabling structured and unstructured financial data to be processed within a unified framework \cite{bommasani2021opportunities}.

The ability to integrate multimodal financial information fundamentally changes how AI interprets financial events. 
Rather than relying solely on isolated numerical variables, modern financial AI combines diverse information sources to capture the broader context surrounding financial decision-making \cite{d2026learning}. 
For example, earnings analysis requires not only structured financial indicators, such as revenue growth and profitability, but also qualitative information from earnings call transcripts and management discussions to understand strategic direction and future expectations \cite{bongale2026comprehensive}. Similarly, market surveillance combines quantitative market indicators with financial news and regulatory updates to distinguish company-specific events from broader market movements, while credit assessment and risk management integrate financial statements with corporate disclosures and market sentiment to identify risks that may not yet be reflected in numerical indicators\cite{ding2015deep}. By integrating these complementary information sources, modern financial AI develops a richer understanding of financial events. This enables AI systems to identify qualitative signals that are difficult to infer from structured numerical data alone and can provide more informed analyses and support better decision-making.

The increasing diversity of financial information also introduces challenges for AI systems. Financial information originates from diverse sources that differ in format, reliability, temporal granularity, and update frequency, making it difficult to construct a consistent representation of evolving financial events and market conditions \cite{pazouki2025artificial}. Effectively integrating these complementary information sources while preserving their semantic meaning and temporal relationships remains an active research challenge. By providing a unified framework for integrating heterogeneous financial information, foundation models enable AI systems to move beyond numerical prediction toward understanding and reasoning over complex financial information, laying the foundation for the intelligent financial tasks discussed in the next subsection.

\subsection{Prediction vs. Intelligent Financial Tasks}
\par \textit{General-Purpose Financial Intelligence:} For decades, financial AI largely followed a task-specific paradigm in which predefined financial variables were transformed into a single prediction or decision. 
This framework proved highly effective for repetitive problems such as stock price forecasting, credit risk estimation, and fraud detection \cite{dixon2020machine}. However, many real-world financial tasks extend far beyond a single prediction. Financial professionals must interpret geopolitical and macroeconomic developments, reconcile information from financial reports and market news, and assess evolving risks under changing market conditions. These tasks require AI systems not only to generate predictions but also to understand financial information, synthesize evidence from multiple sources, reason over complex relationships, and communicate analytical insights. This evolution transformed AI from specialized prediction to general-purpose systems capable of understanding and reasoning over diverse financial information. Leveraging advances in \ac{dl}, foundation models enable a single pretrained model to support financial tasks instead of a single specialized prediction\cite{zhao2024revolutionizing}.

\par \textit{Financial Understanding:}
Traditional financial AI primarily processed unstructured financial information through specialized information extraction tasks. Early systems relied on handcrafted lexicons and rule-based methods to quantify sentiment or identify predefined entities and events, while later \ac{ml} models improved these capabilities through supervised classification \cite{dixon2020machine}. Although effective for narrowly defined objectives such as  sentiment analysis and document classification, these approaches treated financial documents as collections of isolated features rather than coherent sources of knowledge. \Ac{dl} introduced richer semantic representations through architectures such as \acp{rnn} and transformers, enabling models to better capture contextual relationships within financial text and substantially improving tasks such as sentiment analysis, document classification, and named entity recognition \cite{huang2020deep}. However, these models remained largely task-specific, requiring separate training or fine-tuning for individual financial applications. Foundation models transformed this paradigm by enabling holistic understanding of long-form financial documents.  Rather than extracting predefined signals alone, they interpret the context of earnings reports, regulatory filings, news, and reports, supporting semantic search, question answering, summarization, and knowledge retrieval \cite{li2023large}. This evolution shifted financial AI from feature extraction toward comprehensive financial understanding.

\par\textit{Financial Reasoning:}
Understanding financial information alone is often insufficient for solving complex financial problems. Financial professionals must integrate evidence from multiple sources, explain market events, assess competing investment opportunities, identify emerging risks, and evaluate the implications of evolving economic conditions. Traditional AI systems addressed these problems through specialized reasoning pipelines or separate predictive and optimization models, each designed for a specific objective. Advances in \ac{dl} enabled richer semantic representations and transfer learning, improving reasoning within individual financial tasks. However, these capabilities remained largely task-specific and difficult to generalize across diverse analytical problems \cite{goodfellow2016deep,vaswani2017attention}.
Foundation models enable contextual reasoning across heterogeneous financial information, allowing AI systems to connect related evidence, generate coherent explanations, evaluate alternative scenarios, and support complex analytical tasks \cite{son2023}. These capabilities allow financial AI not only to identify patterns but also to interpret their significance, explain analytical conclusions, and support informed financial decision-making.

\par \textit{Integrated Financial Workflows:} As financial AI becomes more capable of understanding and reasoning, analytical tasks become increasingly interconnected within financial workflows. Traditional financial AI decomposed complex analyses into isolated tasks, each addressed by a dedicated model or software component for information retrieval, sentiment analysis, prediction, report generation, or risk assessment. As \ac{dl} matured, individual models became capable of performing increasingly sophisticated functions within their respective domains, yet complex analyses still required manually coordinating multiple independent systems \cite{goodfellow2016deep}. As financial AI acquires stronger understanding and reasoning capabilities, analytical tasks become increasingly integrated into end-to-end financial workflows. This enables AI to retrieve relevant information, interpret financial documents, synthesize evidence, generate analytical reports, and respond to user queries as part of a coherent end-to-end workflow. Consequently, financial AI evolves from a collection of isolated analytical tools into an intelligent financial assistant capable of supporting diverse financial activities through natural language interaction. These integrated workflows naturally extend to retrieval-augmented and agentic AI systems, where retrieval, reasoning, tool use, and content generation are coordinated within multi-step financial processes\cite{d2026learning}.

\subsection{Prediction vs. Financial Reasoning}
While the previous subsection focused on the intelligent tasks enabled by foundation models, their broader impact lies in transforming financial decision-making. Financial decision-making has traditionally relied on the expertise of analysts, portfolio managers, risk officers, and other financial professionals who synthesize quantitative and qualitative information before reaching a decision. Although predictive AI models have become valuable tools for generating forecasts, risk estimates, and classification scores, these outputs represent only one component of a much broader decision-making process. Human experts must still interpret predictions, evaluate supporting evidence, incorporate newly available information, compare alternative courses of action, and exercise professional judgment before acting \cite{cao2022ai}.

Earlier \ac{ml} and \ac{dl} models primarily supported financial decisions by producing increasingly accurate forecasts, risk scores, and classification outputs. While these predictions improved quantitative decision making, they offered limited explanation of why a recommendation was made or how multiple sources of financial information contributed to the result. Earlier \ac{ml} and \ac{dl} models primarily supported financial decisions by generating increasingly accurate forecasts, risk scores, and classification outputs \cite{pazouki2025artificial}. While these predictions improved quantitative decision-making, they provided limited explanation of why a recommendation was made or how different sources of financial information contributed to the result. Foundation models expand the role of AI from generating predictive signals to actively supporting the decision-making process. Rather than relying solely on numerical forecasts, they integrate structured financial indicators with financial reports, market news, analyst reports, and other contextual information to construct a broader view of the decision problem. By synthesizing evidence across these heterogeneous sources, foundation models identify key drivers, evaluate alternative scenarios, and explain the rationale behind their recommendations. This evidence-based approach enables financial professionals to make decisions while improving transparency and explainability of AI-assisted decision support \cite{brown2020language}.

Financial decision-making often occurs in rapidly changing environments where market conditions, corporate disclosures, macroeconomic developments, and regulatory changes continuously reshape the available evidence. Traditional decision support systems typically relied on static datasets, requiring analysts to rerun models whenever new information became available. While foundation models provide natural-language reasoning and explanation, their knowledge remains limited to information available during pretraining. \ac{rag} \cite{lewis2020retrieval} extends this paradigm by grounding financial decision support in continuously updated external knowledge, including market news, regulatory announcements, corporate disclosures, and other real-time information sources.
Rather than relying solely on pretrained knowledge, \ac{rag}-based systems dynamically reassess prior conclusions, explain how new developments influence ongoing decisions, and update recommendations as financial conditions evolve. This evidence-grounded approach enables financial professionals to respond more effectively to changing markets while maintaining human oversight over critical decisions, naturally leading to the collaborative financial intelligence. 

The emergence of agentic AI further extends financial AI from generating recommendations to assisting with multi-step financial decision processes. These systems can continuously monitor evolving market conditions, evaluate alternative courses of action, and coordinate complex analytical workflows. Despite these advances, financial decisions remain high-stakes and subject to regulatory, ethical, and organizational constraints. Consequently, AI is expected to augment rather than replace human judgment, with financial professionals retaining responsibility for validating recommendations and making final decisions. This evolving relationship between AI and human expertise motivates the collaborative paradigms discussed in the following subsection.

\subsection{Automation vs. Collaborative Intelligence}
The evolution of financial AI is transforming how financial work is performed. Rather than supporting isolated analytical tasks, AI is increasingly embedded throughout financial workflows, assisting professionals in research, risk assessment, compliance, reporting, and customer advisory activities. As these systems become integrated into everyday financial practice, the relationship between AI and financial professionals shifts from task automation to sustained collaboration. This transformation reshapes both financial work and the role of expertise across financial organizations \cite{aldridge2026agentic}.

Earlier statistical and \ac{ml} systems primarily supported financial work by automating narrowly defined tasks, such as generating a credit score or flagging a suspicious transaction. These outputs were passed to human professionals with little accompanying context, requiring analysts to interpret the results, gather supporting evidence, and incorporate them into broader decisions. \Ac{dl} improved the accuracy and scope of these automated tasks but preserved the same fundamental relationship: AI generated isolated predictions, while humans remained responsible for interpretation and synthesis. This separation was reinforced by traditional financial organizations, where specialized teams operated sequentially through research, analysis, risk evaluation, compliance, and reporting, creating duplicated effort and decision delays.

Foundation models fundamentally changed this relationship by enabling natural-language interaction between AI and financial professionals. Rather than producing an isolated score or classification, they can explain their reasoning, answer follow-up questions, and synthesize evidence across organizational functions through a shared conversational interface \cite{asaad2026regime}. Retrieval-augmented systems further strengthen this collaboration by grounding AI-generated analyses in external sources, allowing professionals to audit the evidence behind recommendations rather than relying solely on model-generated responses \cite{lewis2020retrieval}. Agentic AI extends this paradigm by taking a more active role throughout the decision-making process: rather than functioning only as analytical tools, agentic systems can retrieve information, coordinate multi-step analyses, and execute portions of financial workflows. As a result, AI becomes an increasingly active collaborator rather than a passive tool \cite{aldridge2026agentic}. 

Increasingly autonomous AI systems introduce new organizational and governance requirements. As AI evolves from decision support to autonomous planning and multi-step task execution, effective human--AI collaboration depends on transparency, accountability, explainability, and clearly defined human responsibilities. Financial institutions must establish governance frameworks that specify when AI recommendations should be accepted, challenged, or overridden while ensuring compliance with regulatory and ethical standards. Consequently, the successful deployment of collaborative financial intelligence depends not only on advances in AI capabilities but also on governance practices that enable responsible and trustworthy human--AI collaboration.

\subsection{Matching AI Paradigms to Financial Tasks}

The paradigms discussed above are often viewed as a strict hierarchy of capability, implicitly suggesting that newer paradigms are always preferable. 
In practice, however, selecting an AI paradigm should depend on the characteristics of the financial task rather than the most advanced capability available. Statistical, \ac{ml}, and \ac{dl} models remain best suited to high-frequency, latency-sensitive prediction tasks with stable underlying relationships. Foundation models become appropriate when a task requires natural-language understanding and reasoning over self-contained information, although their knowledge remains fixed at pretraining. \ac{rag} is needed when decisions depend on continuously evolving information that must be retrieved from verifiable external sources. Agentic AI is appropriate for tasks that coordinate retrieval, reasoning, and tool use across multi-step financial workflows. Table~\ref{tab:ai_paradigm_selection} summarizes these distinctions across six dimensions.
\begin{table*}[t]
\centering
\caption{Guidelines for selecting an appropriate AI paradigm for financial applications. }
\vspace{-.2cm}
\label{tab:ai_paradigm_selection}
\small
\renewcommand{\arraystretch}{1.1}
\begin{tabular}{
>{\raggedright\arraybackslash}p{2.8cm}
>{\raggedright\arraybackslash}p{3.1cm}
>{\raggedright\arraybackslash}p{3.6cm}
>{\raggedright\arraybackslash}p{3.5cm}
>{\raggedright\arraybackslash}p{3.3cm}
}
\toprule
\textbf{Task Characteristic} &
\textbf{Statistical | ML | DL} &
\textbf{Foundation Models} &
\textbf{RAG} &
\textbf{Agentic AI} \\
\midrule
\textbf{Task frequency and latency} &
Best when the task is high-frequency and latency-sensitive &
Suitable for moderate-frequency, non-real-time analysis &
Suitable when retrieval latency is acceptable and frequency is lower &
Reserved for low-frequency, high-value workflows \\
\midrule
\textbf{Knowledge volatility} &
Suitable for stable data relationships over time &
Suitable when answers need no information beyond pretraining &
Necessary when knowledge changes over time &
Necessary for dynamic actions\\
\midrule
\textbf{Reasoning requirement} &
Sufficient for prediction or classification alone &
Sufficient for reasoning over a single document &
Necessary for reasoning across multiple external sources &
Necessary when reasoning spans multi-step planning \\
\midrule
\textbf{Workflow complexity} &
Requires a single prediction &
Requires multiple analytical steps &
Requires retrieval before reasoning &
Requires an end-to-end workflow \\
\midrule
\textbf{Need for traceable evidence} &
A numeric score is sufficient justification &
A natural-language explanation is sufficient &
An evidence-backed, citable recommendation is required &
An evidence-backed execution history is required \\
\midrule
\textbf{Human involvement} &
Human interprets outputs &
AI assists; human judges &
Human validates evidence &
Human supervises actions \\
\bottomrule
\end{tabular}
\end{table*}

For example, summarizing a single earnings call requires neither retrieval nor autonomous coordination, whereas answering a question about current regulatory requirements naturally requires \ac{rag}. In contrast, a full credit underwriting review benefits from agentic AI because it requires coordinating multiple retrieval and analytical steps. A recurring pattern in financial AI deployment is adopting the most capable paradigm regardless of task fit, applying foundation models, \ac{rag}, or agentic AI to narrow, stable, and high-volume problems where the added capability is unnecessary and uneconomical. Effective adoption depends not on the most advanced paradigm available, but on selecting the one that best matches the characteristics of the task.

\section{Open Challenges in Financial AI}
The evolution of financial AI from predictive models to foundation models, retrieval-augmented systems, and agentic AI introduces new challenges that must be addressed before these technologies can be deployed reliably at scale. Although recent advances have substantially expanded AI capabilities, realizing their full potential requires overcoming challenges related to reasoning, knowledge integration, evaluation, governance, and human--AI collaboration.

\paragraph{Financial Data Readiness}
Financial AI increasingly relies on integrating structured and unstructured information from diverse sources. However, much of this information remains fragmented across proprietary systems, legacy infrastructures, and organizational data silos, making reliable access, integration, and governance difficult. As foundation models, retrieval-augmented systems, and agentic AI become increasingly dependent on timely and multimodal data, ensuring the quality, and reliability of external information becomes equally important. Developing AI-ready financial data infrastructures that support secure, trustworthy, and continuously updated data access remains a fundamental challenge.

\paragraph{Interpretability and Explainability}
Traditional financial AI relied on transparent statistical models and manually designed decision rules, making the reasoning process relatively easy to inspect. As financial AI evolved toward autonomous systems, predictive performance has improved substantially, but transparency has declined. Although these systems can generate natural-language explanations, whether these faithfully reflect the underlying reasoning—rather than plausible post-hoc narratives—remains an open question, particularly for high-stakes financial decisions.

\paragraph{Evaluation and Benchmarking}
Traditional financial AI was primarily evaluated using well-defined predictive metrics, such as accuracy, forecasting error, or trading performance. As financial AI evolves from prediction to reasoning and autonomous decision-making, these metrics are no longer sufficient. Modern financial AI also requires evaluating reasoning, retrieval quality, decision making, and alignment with regulatory objectives. Developing standardized, finance-specific benchmarks and reproducible evaluation frameworks remains an important research challenge.

\paragraph{Governance of Autonomous Systems}
Traditional financial governance assumes that decisions are made by identifiable humans. As financial AI evolves from decision support to autonomous system, this assumption no longer holds. Determining responsibility, ensuring regulatory compliance, and maintaining effective human oversight become increasingly difficult as autonomous agents participate in financial decision-making. Moreover, interactions among multiple autonomous agents may introduce unforeseen operational and systemic risks, requiring new governance frameworks that clearly define accountability between humans and AI systems.


\section{Conclusions}
This survey presented a unified perspective on the evolution of AI in finance, tracing its progression from statistical learning and \ac{ml} to \ac{dl}, foundation models, \ac{rag} and agentic AI. Rather than viewing these paradigms as successive replacements, we argued that they provide complementary capabilities suited to different classes of financial tasks. We further showed how this evolution has transformed financial information, intelligent financial tasks, decision support, and human--AI collaboration, and proposed practical guidelines for matching AI paradigms to task characteristics. As financial AI continues to evolve toward increasingly autonomous systems, future progress will depend not only on more capable models but also on reliable data, trustworthy reasoning, robust evaluation, and effective governance.
\vspace{-.2cm}
\bibliographystyle{ACM-Reference-Format}
\bibliography{sample-base}

\appendix

\end{document}